\documentclass[sigconf]{acmart}

\setcopyright{none}
\renewcommand\footnotetextcopyrightpermission[1]{}

\usepackage{url}
\usepackage{fancyhdr}
\fancypagestyle{plain}{%
   \fancyhf{} %
}
\usepackage{balance}

\begin{document}

\title{Proof of Steak}

\author{Jon Crowcroft}
\affiliation{
	\institution{University of Cambridge}
}

\author{Hamed Haddadi}
\affiliation{
	\institution{Imperial College London}
}

\author{Arthur Gervais}
\affiliation{
	\institution{Imperial College London}
}

\author{Tristan Henderson}
\affiliation{
	\institution{University of St Andrews}
}

%

\begin{abstract}
We introduce \emph{Proof-of-Steak} (PoS) as a fundamental net-zero block generation technique, often accompanied by Non-Frangipane Tokens. Genesis cut is gradually heated and minted (using the appropriate sauce), enabling the miners to redirect the extracted gold and the dissipated heat into the furnace, hence enabling the first fully-circular economy ever built using blockchain technology, utilising tamper-evident steak hach\'{e}. In this paper we present the basic ingredients for building Proof-of-Steak, assessing its global impact, and opportunities to save the world and beyond!
\end{abstract}

\ccsdesc[500]{Networks~Routing protocols}
\ccsdesc[500]{Security and privacy~Security protocols}

\keywords{Steak, NFT, Blockchain}

\maketitle

\section{Introduction to Proof-of-Steak}\label{sec:intro}

To ensure the success of our new chain of high end steak restaurants, it is paramount to ensure trust in our product. People need to know that when they get an Angus Beef Sirloin, or a Fiorentino Porterhouse, or a Kobe Wagu Tornados, you are getting the genuine article, not some dodgy knockoff from England.\footnote{\url{https://www.forbes.com/sites/larryolmsted/2012/04/12/foods-biggest-scam-the-great-kobe-beef-lie/}}

To this end, we have implemented the world's first \emph{Proof-of-Steak} technology, based on the famous bloodline going back decades to early systems like the Internet's Certificate Transparent Authority, and Notarized smart contracts from the Nakatomi corporation. Don't be put off by gossip about our earlier technology, where we used to gather and weigh the sweat from the masseurs that look after the living cattle so well, and use this as a proxy for proof of work. Of course, an underground market in fake "Kobe sweat" quickly arose, mainly centred around gyms and boxing rings, where there was ample supply of people desperate for a quick free energy drink (ironically, given the main typical ingredient is Taurine). Now, as with many other such setbacks in the successful commercial world, we have moved on, embracing the new way to assure all steps along the supply chain with the brand new \emph{Proof-of-Steak}\texttrademark.

When you see this brand (literally, on the animal, and on your plate) you can be assured that at every stage, a second creature or its derivative, have been securely sequestered in a third party store, ahead of time, escrowed in case any party along the line has been tempted to stray, or indeed third party rustlers (as they are still known) have replaced the genuine article with counterfeit products.

We have tested these systems thoroughly, recently with a high profile pen tester black-and-white team from the cattle ranchers' institute, who were able to show that gold leaf used in one particular product line was only 23-carat, although all the other ingredients were worth their salt.

While Miner Extractable Value is threatening the steak's incentive stability, we propose a novel mechanism to atomically extract and bind MEV to the cows' emitted methane and resulting carbon dioxide. Our cows are therefore, to the best of our knowledge, the first cows to not feel guilty for producing steak.

\begin{figure}[t]
\begin{center}
\includegraphics[width=\linewidth]{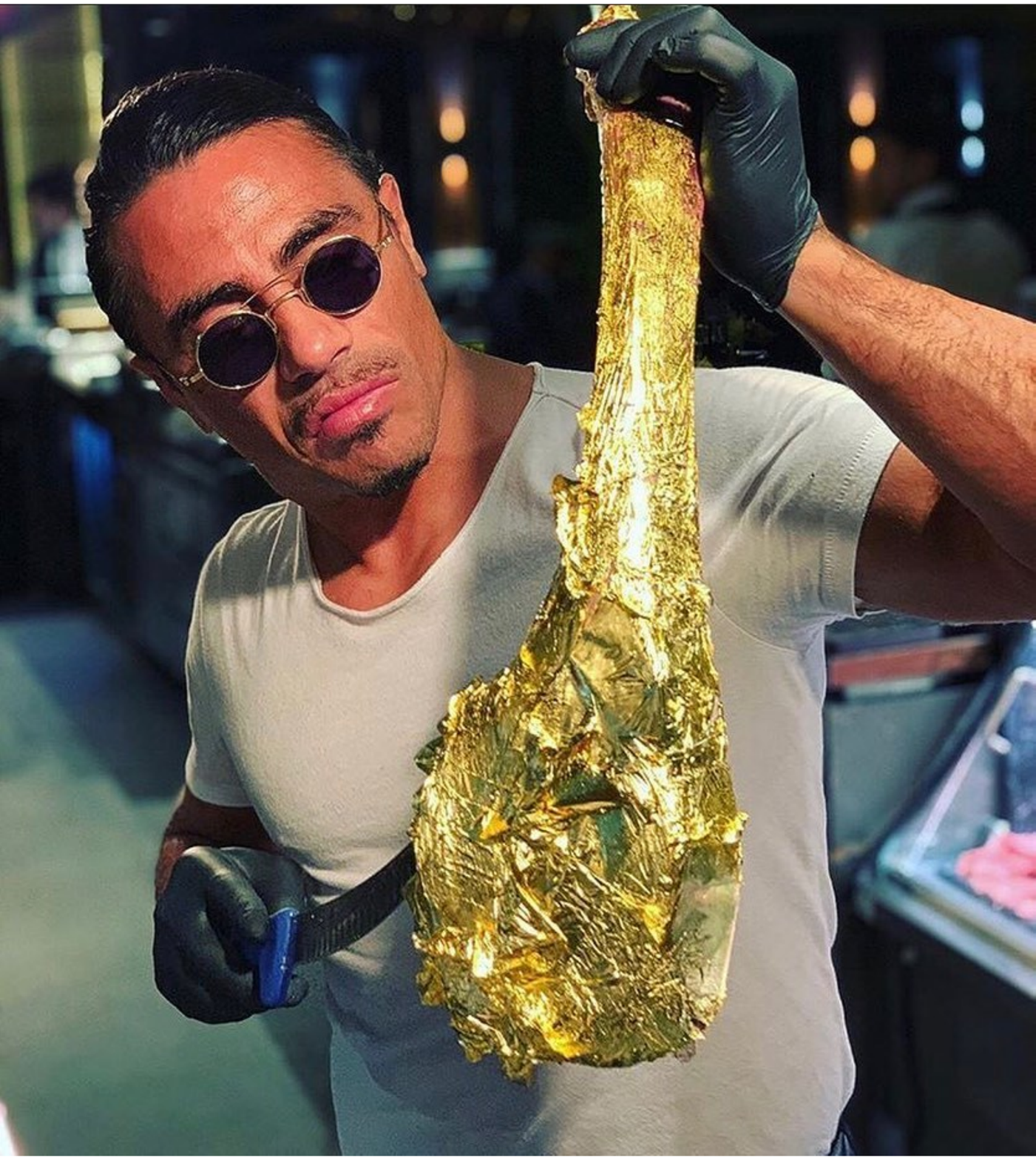}
\caption{Minted golden steak \footnotesize{(Source: The Sun)}}
\label{default}
\end{center}
\end{figure}

\subsection{An Alternative to Proof-of-Wok}
There has been centuries of progress in an alternative block preparation technique, \textit{Proof-of-Wok} (PoW), where a pre-heated wok is used to mint the steak. However, uncertainties on the burn-rate of wok, its half life, and the recent crackdowns in China on crypto mining have all prevented a stable supply of woks for distributed and decentralised restaurants. We leave the ideal sharding strategies and energy comparisons of PoS and PoW as future work.

\section{Animal Rights}
Animal welfare and rights are also crucial for creating a trustworthy and reliable Proof-of-Steak platform. We go beyond traditional animal welfare requirements by affording our animals all of the rights granted to ``natural persons'' under the GDPR. To comply with GDPR Article $25$ on data protection by design, we use differential privacy and distributed analytics~\cite{10.1145/2185376.2185390} to obfuscate the identity of any given Kobe cow so as to prevent them from being singled out. It goes without saying that all passwords in our system are salted. In keeping with GDPR Article 6 on the lawfulness of processing, we are careful to only process animals when it is in our legitimate interest to turn them into tasty, overpriced, social-media-friendly morsels.

To exceed regulatory requirements by $42$ orders of magnitude for protecting animal rights, we show that differential privacy coupled with privacy preserving federated learning and optional trusted hardware enclaves is the long awaited secret sauce.

\section{Related Work}

Earlier work in the EU on an exchange rate mechanism for wine and dairy herd products fell foul of a badly designed Proof-of-Snake protocol, which used cutting-edge mobile-phone gaming technology to attempt to bound the length of any ledger by linking the head and tail of the chain, and quite literally, devour the body to prevent it growing without bound. Insufficient attention was paid to the dynamics of various parties' periodic re-skinning of the scheme, so that stability was never assured. Not well done. The problems with deep fakes are outside the scope of this cuisine.\footnote{\url{https://www.manchestereveningnews.co.uk/whats-on/food-drink-news/man-recreates-salt-baes-1500-21958035}}

\section{Future work}
There is so much more to discuss: nothing at steak, weak subjectivity, slashing steak, steaking rewards, time to finality, and so on. We will also leave the attacks on PoW such as apeing in and getting rekt to future works (and cooler woks). We intend to integrate the proof of steak with the new Cambridge Carbon Credit Certification system, by measuring the herd methane emissions, so that customers who wish to pepper their steak decisions with green sauce will have that option, although we expect this to be rare. We note with respect their use of mether, which replaces the nearest rival's ether with 50\% lower carbon. An important block in the PoS system is the full integration of the user output into the Ambient Loo~\cite{6dfa4128c7b44fa3a18757a23ca4c30b} towards a circular carbon-neutral economy, which we will leave as an exercise for the green computing and the chemical engineering communities. Moreover, while our methane-MEV binding process is the first of its kind, we hope that future work can increase its efficiency by at least $1337\times$. We follow the community consensus in discounting the Maillard reaction, due to its inherent nucleophilicity.

We believe that this is the first successful application of cyber-physical supply chain assurance to use proof-of-steak, but we do not believe it will be the last. The timing could not be better, as we meet all of the United Nations' Sustainable Development Goals, just when global heating could rise to the point where the origin herds are no longer viable. We are part of the future. Please join us for a fine meal. We shall source excellent beverages and carbon offset all of our cooking technology and practices with certificated reforestation programs, to the point. Large-scale adoption can enable us to feed the world, and let them know it's Christmas time,\footnote{\url{https://www.youtube.com/watch?v=cIxj7Ew_99w}} one steak at a time. 

\begin{acks}
We wish to thank the various field experts who strongly supported us and provided extensive feedback, but could not be named due to the ongoing conflict of interests in the cryptocurrency or the catering industry.
\end{acks}

{
{
	\bibliographystyle{ACM-Reference-Format}
	\bibliography{literature}


\begin{thebibliography}{2}


\ifx \showCODEN    \undefined \def \showCODEN     #1{\unskip}     \fi
\ifx \showDOI      \undefined \def \showDOI       #1{#1}\fi
\ifx \showISBNx    \undefined \def \showISBNx     #1{\unskip}     \fi
\ifx \showISBNxiii \undefined \def \showISBNxiii  #1{\unskip}     \fi
\ifx \showISSN     \undefined \def \showISSN      #1{\unskip}     \fi
\ifx \showLCCN     \undefined \def \showLCCN      #1{\unskip}     \fi
\ifx \shownote     \undefined \def \shownote      #1{#1}          \fi
\ifx \showarticletitle \undefined \def \showarticletitle #1{#1}   \fi
\ifx \showURL      \undefined \def \showURL       {\relax}        \fi
\providecommand\bibfield[2]{#2}
\providecommand\bibinfo[2]{#2}
\providecommand\natexlab[1]{#1}
\providecommand\showeprint[2][]{arXiv:#2}

\bibitem[\protect\citeauthoryear{Haddadi, Henderson, and Crowcroft}{Haddadi
  et~al\mbox{.}}{2010}]%
        {6dfa4128c7b44fa3a18757a23ca4c30b}
\bibfield{author}{\bibinfo{person}{Hamed Haddadi}, \bibinfo{person}{Tristan
  Henderson}, {and} \bibinfo{person}{Jon Crowcroft}.}
  \bibinfo{year}{2010}\natexlab{}.
\newblock \showarticletitle{The ambient loo - caught short when nature calls?}
\newblock \bibinfo{journal}{\emph{ACM Computer Communication Review}}
  \bibinfo{volume}{40}, \bibinfo{number}{2} (\bibinfo{date}{April}
  \bibinfo{year}{2010}), \bibinfo{pages}{78--78}.
\newblock
\showISSN{0146-4833}
\newblock
\shownote{This article is an editorial note submitted to CCR. It has not been
  peer reviewed. The author takes full responsibility for this
  article{\textquoteright}s technical content. Comments can be posted through
  CCR Online.}


\bibitem[\protect\citeauthoryear{Haddadi, Mortier, Hand, Brown, Yoneki,
  McAuley, and Crowcroft}{Haddadi et~al\mbox{.}}{2012}]%
        {10.1145/2185376.2185390}
\bibfield{author}{\bibinfo{person}{Hamed Haddadi}, \bibinfo{person}{Richard
  Mortier}, \bibinfo{person}{Steven Hand}, \bibinfo{person}{Ian Brown},
  \bibinfo{person}{Eiko Yoneki}, \bibinfo{person}{Derek McAuley}, {and}
  \bibinfo{person}{Jon Crowcroft}.} \bibinfo{year}{2012}\natexlab{}.
\newblock \showarticletitle{Privacy Analytics}.
\newblock \bibinfo{journal}{\emph{SIGCOMM Comput. Commun. Rev.}}
  \bibinfo{volume}{42}, \bibinfo{number}{2} (\bibinfo{date}{mar}
  \bibinfo{year}{2012}), \bibinfo{pages}{94–98}.
\newblock
\showISSN{0146-4833}
\urldef\tempurl%
\url{https://doi.org/10.1145/2185376.2185390}
\showDOI{\tempurl}


\end{thebibliography}
}
}

\end{document}